\documentclass[sigplan,nonacm]{acmart}
\usepackage[utf8]{inputenc}
\usepackage{natbib}
\usepackage{float}
\usepackage{caption}
\usepackage{algorithm}
\usepackage{algpseudocode}
\usepackage{listings}
\usepackage[capitalize]{cleveref}
\usepackage{booktabs}
\usepackage{tabularx}
\usepackage{graphicx}
\usepackage{xcolor}
\usepackage{multicol}

\title{NLP-Based .NET CLR Event Logs Analyzer}

\author{Maxim Stavtsev}
\email{mastavtsev@edu.hse.ru}
\orcid{0009-0008-8171-8167}
\affiliation{\institution{HSE University}\city{Moscow}\country{Russia}}

\author{Sergey Shershakov}
\email{sshershakov@hse.ru}
\orcid{0000-0001-8173-5970}
\affiliation{\institution{HSE University}\city{Moscow}\country{Russia}}

\begin{document}

\begin{abstract}
In this paper, we present a tool for analyzing .NET CLR event logs based on a novel method inspired by Natural Language Processing (NLP) approach. Our research addresses the growing need for effective monitoring and optimization of software systems through detailed event log analysis. We utilize a BERT-based architecture with an enhanced tokenization process customized to event logs. The tool, developed using \textsc{Python}, its libraries, and an \textsc{SQLite} database, allows both conducting experiments for academic purposes and efficiently solving industry-emerging tasks. Our experiments demonstrate the efficacy of our approach in compressing event sequences, detecting recurring patterns, and identifying anomalies. The trained model shows promising results, with a high accuracy rate in anomaly detection, which demonstrates the potential of NLP methods to improve the reliability and stability of software systems.

\noindent
\textbf{Demo video:} \href{https://youtu.be/JLCS4F-AlYc}{YouTube} \\
\textbf{GitHub:} \href{https://github.com/mastavtsev/NLP-CLR-LogAnalyzer}{NLP-CLR-LogAnalyzer}

\end{abstract}

\maketitle

\section{Introduction}
Most organizations use various software systems, necessitating effective monitoring and resource allocation. Event logs, as primary artifacts of software operations, are crucial to understanding system functions and identifying optimization opportunities. Process mining combines process science and data analysis methods to extract value from such logs, which allows the optimization of processes. This project extends the approach proposed by Stepanov and Mitsyuk \cite{stepanov2024extracting} by enhancing low-level .NET event log analysis in two ways: 1) pattern detection, to understand system interactions, and 2) anomaly detection, to identify and prevent abnormal behaviors.

We apply neural network models for automated and scalable analysis. Unsupervised learning is utilized due to large, unlabeled datasets typical in software systems. Using transformer-based NLP methods, we tokenize event traces to identify patterns and anomalies. This work demonstrates the effective application of NLP techniques to .NET CLR event log analysis.

This article is organized as follows: 
\cref{sec:related}  provides an overview of existing solutions for event log analysis, 
\cref{sec:algorithms} describes the algorithms used in the project, 
\cref{sec:method} presents the proposed method for event log analysis, including training the machine learning model and implementation of the proposed algorithms, and finally
\cref{sec:conclusion} offers a summary of the article.

\section{Related Work}\label{sec:related}

\begin{figure}
  \centering
  \includegraphics[scale=0.3]{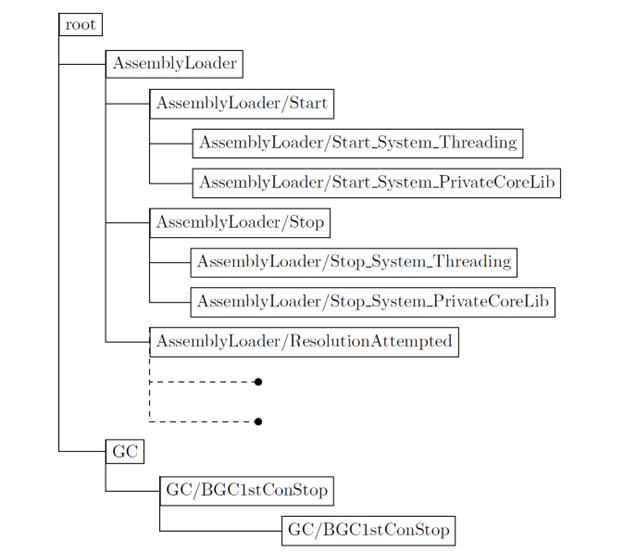}
  \caption{An example of a predefined hierarchy used to raise the abstraction level of low-level event logs, where "root" is the most abstract event and \textit{AssemblyLoader/Start\_System\_Threading} is the most specific, detailed event.}
  \Description{}
  \label{fig:hierarchy}
\end{figure}

In \cite{stepanov2024extracting} authors propose a method for extracting high-level activities from low-level event logs of program execution. To achieve this, they developed a tool called \textsc{Procfiler}, which collects events that occur during the execution of programs written in the C\# programming language in the .NET CLR runtime environment and creates a log from them. They then apply a predefined hierarchy to raise the abstraction level of the events. An example of such a hierarchy is shown in \Cref{fig:hierarchy}. The root is an artificially added, most abstract event. For instance, \textit{AssemblyLoader/Start\_System\_Threading} represents a very specific and detailed event. This event can be abstracted to a higher level by naming it \textit{AssemblyLoader/Start}.

In our project, we utilized the results of the Procfiler tool, specifically logs with the lowest level of abstraction, meaning they contain events that are the leaves in the hierarchical tree shown in Figure 1.

For the task of anomaly detection in event logs, supervised learning methods have been applied, treating this task as a binary classification problem \cite{huang2020hitanomaly}. However, this significantly reduces the applicability of such methods in real systems, as it requires a prelabeled dataset, and more importantly, limits the ability to detect previously unseen anomalies. There are works that use unsupervised learning approaches \cite{du2017deeplog} based on LSTM, as well as the \textsc{BERT} model \cite{lee2023lanobert}, which is based on the transformer architecture and employs preliminary tokenization of the event log.

To the best of our knowledge, no previous study has investigated the applicability of NLP methods for automated pattern and anomaly detection for the domain of low-level event .NET CLR logs.

\section{Algorithms}\label{sec:algorithms}
There are some major limitations to the rapid and efficient analysis by process mining methods. Among them are the large volume of event logs and the need to perform a preliminary analysis of the input data to understand the structure of interacting process elements. One of the most important hypotheses in this work is that approaches adapted from the field of NLP can remove such limitations.

\subsection{Event Log Encoding}\label{sec:encoding}

Most NLP algorithms are applied to sequential data. For the task of event log analysis, we need to represent logs as sequences for further application of the algorithm.

\Cref{tab:event_log} shows an example fragment of an event log, where each event has three mandatory attributes: it is the realization of some activity from the set of activities $\mathbb{A}$ — $act_i$, a timestamp $ts_i$ corresponding to the event's start time, and a process identifier during which it was executed. The set $\mathbb{A}$ is finite and defined by the set of activities allowed in the .NET runtime environment, their list is presented in the Appendix of the work \cite{stepanov2024extracting}.

\begin{table}[ht]
    \centering
    \caption{Example fragment of a low-level event log from the CLR environment.}
    \small 
    \begin{tabular}{|c|p{4.5cm}|c|}
        \hline
        Trace ID & Activity & Time \\ \hline
        31237 & \textit{Method/MemoryAllocatedForJitCode} & 14:27:57 \\ \hline
        -1 & \textit{Method/LoadVerbose} & 14:27:57 \\ \hline
        31237 & \textit{GC/SampledObjectAllocation} & 14:27:57 \\ \hline
        31237 & Buffer/Returned & 15:37:34 \\ \hline
        \ldots & \ldots & \ldots \\ \hline
    \end{tabular}
    \label{tab:event_log}
\end{table}

The considered event log contains events related to the .NET application's execution thread (e.g. ID 31237) and system threads (e.g., ID -1). These events can be combined into one trace by merging events according to their IDs and timestamps, resulting in the trace $trace_1 = \langle \text{Method/MemoryAllocatedForJitCode}, \\ \allowbreak
\text{GC/SampledObjectAllocation}, \allowbreak
\text{Method/LoadVerbose}, \\ \allowbreak
\text{Buffer/Returned} \rangle$.

By using this approach, we obtained the final set of event log traces.

In order to represent traces as textual sequences, each activity from the set of all allowed activities $\mathbb{A}$ in this work is encoded with a unique non-control Unicode character\footnote{\url{https://en.wikipedia.org/wiki/List_of_Unicode_characters}}, let $\mathbb{U}$ be the subset of these characters. Thus, we formed a bijection between the set $\mathbb{A}$ and the set $\mathbb{U}$,
$f: \mathbb{A} \longleftrightarrow \mathbb{U}$.

For example, consider a bijective function $f_0 \subset f$, defined by the set of pairs.

\begin{equation*}
\begin{aligned}
f_0 = \{ & (\text{Method/MemoryAllocatedForJitCode}, "a"), \\
      & (\text{Method/LoadVerbose}, "b"), \\
      & (\text{GC/SampledObjectAllocation}, "c"), \\
      & (\text{Buffer/Returned}, "d") \}
\end{aligned}
\end{equation*}

Then the trace $trace_1$ can be represented as the sequence $seq_1 = "acbd"$.

We reduced the task of representing a trace to a sequence of Unicode characters, solvable using NLP tokenization algorithms such as \textsc{BPE}, \textsc{WordPiece}, and \textsc{Unigram}. We chose the \textsc{BPE} algorithm because it preserves a dictionary of tokens, merging the most common pairs. After completing the tokenizer training, we obtained the final set of permitted tokens, denoted as $\mathbb{T}$, where each token represents a subsequence of Unicode characters.

Any sequence $seq_i$ can be represented as $tokens_i = (t_j: t_j \in \mathbb{T}, i = 1, \ldots, \varphi(seq))$, where $\varphi(seq)$ determines the number of tokens. The tokenization process, defined as $\mathcal{T}: seq_i \rightarrow tokens_i$, converts $seq_1$ into the set $tokens_1 = ['ac', 'bd']$. Tokenization also compresses traces, reducing the input sequence length significantly. 

Each token is encoded with a unique number corresponding to a value in the embedding table (numeric vectors), which in turn are trainable parameters of the neural network, the configuration of which we will describe in \Cref{sec:network}. Using numeric vectors, we can encode traces and feed them to the neural network input.

\subsection{Neural Network Configuration}\label{sec:network}

A key algorithm in deep learning, especially in NLP, is the transformer architecture \cite{vaswani2017attention}, which consists of an encoder and a decoder. The encoder processes the input sequence with the attention mechanism and feed-forward layers, producing vectors that the decoder further transforms into a probability vector. \textsc{BERT} \cite{devlin2018bert} is a transformer model that is based only on the encoder and applies attention to tokens based on their context within a sequence.

One of the ways to train \textsc{BERT} is the Masked Language Modeling (MLM) approach, which involves masking a certain percentage of randomly selected tokens with a special token [MASK]. During training, the model aims to minimize the loss function's error by predicting the token hidden behind the [MASK]. We claim that a model trained on unlabeled correct (without anomalies) event logs can detect events that do not match the context of normal behavior. For this reason, in this work, we apply a BERT-based model for anomaly detection.

After analyzing existing BERT-based model architectures, we selected the \textsc{SqueezeBERT} architecture \cite{iandola2020squeezebert} for this work. The authors of the work presenting this architecture show that the majority of the model's parameters and the bulk of the time during its application are concentrated in the feed-forward layers. They propose a convolution-based approach borrowed from the field of computer vision to optimize the neural network. As a result, they manage to reduce the number of parameters to approximately 40 million, compared to 100 million in original \textsc{BERT}, which requires less computational resources, without significant loss of quality.

\section{Method}\label{sec:method}
\subsection{Patterns Detection}

\begin{algorithm}
\scriptsize
\caption{Algorithm for Extracting Traces}
\label{alg:extract_traces}
\begin{algorithmic}[1]
\Function{ProcessXESTraces}{input\_filepath}
    \State $log \gets ReadXES(input\_filepath)$
    \State $event\_log \gets FilterLogForNeededColumns(log)$
    \State $needed\_indexes \gets GetNeededIndexes(event\_log)$
    \State $outliers \gets IdentifyOutliers(event\_log, needed\_indexes)$
    \State $event\_log \gets FilterLogByIndexes(event\_log, needed\_indexes)$
    \State $traces\_log \gets CreateTracesLog(event\_log)$
    \State $final\_trace\_log \gets IntegrateOutliersIntoTraces(traces\_log, outliers)$
    \State $list\_of\_traces \gets ConvertTracesToList(final\_trace\_log)$
    \State \Return $list\_of\_traces$
\EndFunction
\end{algorithmic}
\end{algorithm}

\begin{algorithm}
\scriptsize
\caption{Algorithm for Tokenizing Traces}
\label{alg:tokenize_traces}
\begin{algorithmic}[1]
\Function{ProcessTracesToSequences}{traces, LoA}
    \State $accepted\_events \gets LoadAcceptedEvents()$
    \State $event\_codes \gets MapEventsToCodes(accepted\_events)$
    \State $sequences \gets []$
    \For{each $trace$ in $traces$}
        \State $sequence \gets ConvertTraceToSequence(trace, event\_codes)$
        \State $sequences.append(sequence)$
    \EndFor
    \State $processed\_traces \gets TokenizeSequences(sequences, LoA)$
    \State \Return $processed\_traces$
\EndFunction
\end{algorithmic}
\end{algorithm}

Tokens in the tokenizer's dictionary reflect frequently occurring interactions, thus considered patterns in this work. For instance, a group of events events encoded by symbols $a$, $b$, $c$ appearing as the token $bac$ in the event trace is a pattern. We trained 13 tokenizers with dictionary sizes from 512 to 20,000 tokens, some with a maximum token length limit, to analyze these patterns at different abstraction levels — higher levels encode larger numbers of events into single tokens.

The pattern detection process involves two algorithms. \Cref{alg:extract_traces} describes obtaining a list of traces from raw CLR low-level event logs by extracting necessary columns and combining events by timestamps. \Cref{alg:tokenize_traces} extracts tokens from event traces at a specified abstraction level (LoA), forming a list of acceptable events and applying a mapping to create sequences. These sequences are then tokenized using the trained tokenizers, resulting in a list of tokens for each trace.

Consider an example of the results of the pattern search algorithms on event logs of 25 C\# program runs. 

\begin{figure}
\includegraphics[scale=0.4]{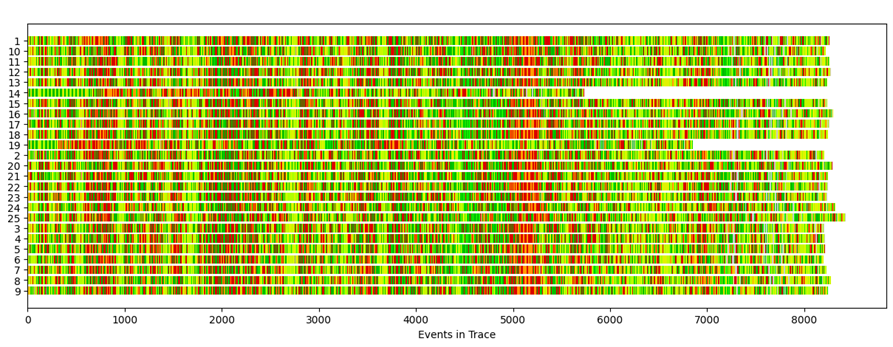}
\caption{Non tokenized log}
\Description{}
\label{fig:non_tokenized_log}
\end{figure}

\begin{figure}
\includegraphics[scale=0.4]{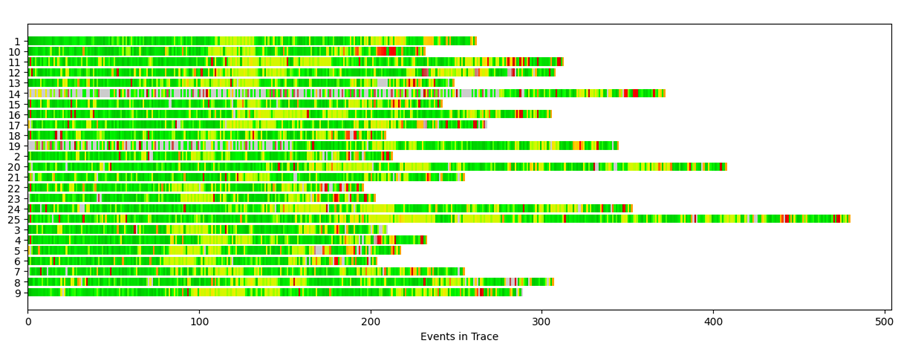}
\caption{Tokenized log with LoA 10}
\Description{}
\label{fig:log_loa_10}
\end{figure}

We visualized the obtained traces, showing the trace ID on the horizontal axis and the number of events (tokens) on the vertical axis, with each color representing a different token/event. Tokenization not only significantly reduced the average trace length from 8000 events in the non-tokenized log (\Cref{fig:non_tokenized_log}) to 200–300 tokens at an abstraction level of 10 (\Cref{fig:log_loa_10}), but also produced a list of frequently co-occurring events during program execution, which is the set of tokens of the encoded trace of program execution. This list can be further analyzed by domain experts to gain deeper insights into the execution process and potentially identify performance bottlenecks.

The method of pattern detection presented in this work is an alternative to the repeated alphabets method presented in \cite{stepanov2024extracting}. Thus, we can say that our method is more versatile as it is used not only for pattern search, but also when using the \textsc{SqueezeBERT} neural network.

\subsection{Anomalies Detection}

Anomaly detection is based on the \textsc{SqueezeBERT} neural network architecture, and there are two main approaches in the NLP field for using machine learning models. The first approach involves fine-tuning an already trained model for specific tasks, which is usually optimal. However, this approach is not feasible in our case as the BERT-based model has not been previously applied to .NET CLR event logs. Therefore, we train the model from scratch using a tokenizer with a maximum abstraction level of 13, a dictionary size of 20,000 tokens, and a maximum token length of 300 characters.

We used the LAMB optimizer \cite{you2019large} for training. The final model consists of 43.6 million parameters and was trained on the same dataset used for tokenizers, employing a tokenizer with a maximum abstraction level of 13, which has a vocabulary size of 20,000 tokens, and a maximum token length of 300 characters. The context window size was set to 512 tokens, with shorter traces padded using the [PAD] token. Training was conducted for 300 epochs in the Google Colab environment on an Nvidia Tesla A100 GPU.

\begin{algorithm}
\scriptsize
\caption{Algorithm for Anomaly Detection}
\label{alg:anomaly_detection}
\begin{algorithmic}[1]
\Function{EvaluateTraces}{traces}
    \For{each trace in traces}
        \State $tokens \gets ProcessTracesToSequences(trace, 13)$
        \State $(probs, loss) \gets EvaluateByTokens(tokens)$
        \State $brier \gets EvaluateTraceBrier(tokens)$
        \State $count\_abnormal \gets 
        CountNonEmpty([probs, loss, brier])$
        \If{$count\_abnormal \geq 2$}
            \State Print "Trace " + trace + " is abnormal."
        \Else
            \State Print "Trace " + trace + " is normal."
        \EndIf
    \EndFor
\EndFunction
\end{algorithmic}
\end{algorithm}

The anomaly detection algorithm is described in \Cref{alg:anomaly_detection}. It takes a list of traces as input, which are subsequently tokenized based on \cref{alg:tokenize_traces}. Then, it performs an evaluation using two methods: probability-based and loss-based, as well as Brier score evaluation.

The probability and loss evaluation is performed by masking each token in a trace with the [MASK] token, applying the \textsc{SqueezeBERT} model, and comparing the model output with the observed value. If the probability of the observed token is less than 0.85, it is considered anomalous. Similarly, if the loss function value is greater than 0.05, the token is considered anomalous.

The Brier score helps detect anomalies at group token levels, increasing the accuracy of detecting incorrect behavior. By masking 20\% of randomly selected tokens and calculating the Brier score for them, we determine if the entire trace is anomalous if the score exceeds 0.5.

We also used a SQLite database to store these three evaluations for each trace. If an identical trace appears again, we retrieve the scores from the database instead of rerunning the model. This approach saves time and computational resources.

\begin{figure}
\includegraphics[scale=0.4]{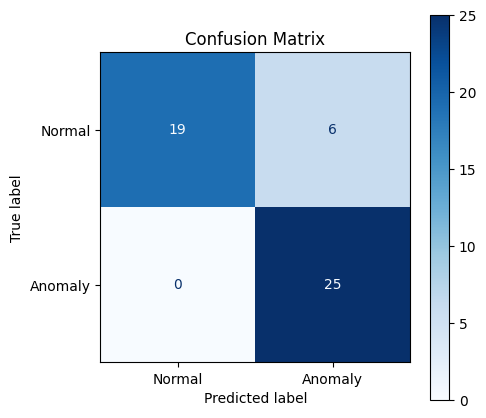}
\caption{Confusion matrix of model performance}
\Description{}
\label{fig:confusion_matrix}
\end{figure}

Model validation was performed on the basis of 25 synthetically generated anomalous traces, as well as 25 traces with normal behavior. Anomalous traces were obtained by adding five random events at random positions, simulating the expected anomalous behavior in the trace. The validation results are presented in the confusion matrix in \Cref{fig:confusion_matrix}.

As observed in \Cref{fig:confusion_matrix}, the developed model shows satisfactory results; however, it makes errors in 6 cases. This error is less problematic because falsely labeling a normal trace as anomalous is preferable to missing an actual anomaly. Increasing the volume of training data could improve the model's quality, so further training on a larger dataset is recommended.

\section{Conclusion}\label{sec:conclusion}

In this study, existing NLP approaches applied to the analysis of event logs were examined. We developed a tool in \textsc{Python}, which is designed for analyzing patterns and anomalies in logs of .NET CLR applications. Our tool supports multiple levels of abstraction for pattern detection and utilizes an \textsc{SQLite} database to store and reference previously analyzed traces, optimizing performance. A model based on the \textsc{BERT} architecture, specifically \textsc{SqueezeBERT}, was trained from scratch. Validation results demonstrate that the model performs well in anomaly detection, and it is expected that increasing the volume of data can improve its quality. Thus, we have shown that NLP approaches can be effectively applied to the analysis of event logs in the .NET CLR runtime environment.

\bibliographystyle{ACM-Reference-Format}
\bibliography{main}

\end{document}